\documentclass[aps,prd,amsmath,amssymb,preprintnumbers,twocolumn]{revtex4}
\usepackage[a4paper,left=20mm,right=20mm,top=25mm,bottom=25mm]{geometry}
\usepackage{amsmath,amssymb,graphicx,color} 
\usepackage{epsf}
\usepackage{slashed}  
\usepackage{cancel}
\usepackage{bbold}
\usepackage{youngtab}
\usepackage{multirow}

\newcommand{\SO}{\text{SO}}
\newcommand{\SU}{\text{SU}}
\newcommand{\U}{\text{U}}
\newcommand{\Sp}{\text{Sp}}

\usepackage[colorlinks,linkcolor=blue,citecolor=blue,urlcolor=blue,linktocpage]{hyperref}
\newcommand{\hhref}[2][]{\href{http://arxiv.org/abs/#2#1}{arXiv:#2}}

\begin{document}

\title{\Large  Light 't~Hooft Top Partners}
\author{Giacomo Cacciapaglia$^a$}
\email{g.cacciapaglia@ipnl.in2p3.fr}
\author{Alberto Parolini$^b$}
\email{parolini85@kias.re.kr} 
\affiliation{$^a$Universit\'e de Lyon, France; Universit\'e Lyon 1, Villeurbanne, France; \\
CNRS/IN2P3, UMR5822, IPNL F-69622 Villeurbanne Cedex, France;}
\affiliation{$^b$Quantum Universe Center, Korea Institute for Advanced Study, Seoul 130-722, Korea.}

\begin{abstract}
Vector-like quarks, usually dubbed top partners, are a common presence in composite Higgs models. Being composite objects, their mass is expected to be of the order of their inverse size, that is the condensation scale of the new strong interactions. Light top partners, while not being a generic prediction, are however often considered in phenomenological models. We suggest that their lightness may be due to the matching of global 't Hooft anomalies of the underlying theory.
We check this mechanism in explicit models showing that, in one case, composite fermions with the quantum numbers of the top quark obtain a mass which is controlled by a soft breaking term and can be made parametrically small.
 \\[2mm]
\end{abstract}

\maketitle


Compositeness is an attractive hypothesis to solve the Standard Model (SM) hierarchy problem. In modern Composite Higgs Models (CHM), the Brout-Englert-Higgs doublet arises as a pseudo Nambu-Goldstone Boson (pNGB) due to a new strongly interacting sector that breaks spontaneously a global symmetry, not shared by the SM fields, to a properly chosen subgroup. Since it is a Goldstone, shift symmetry forces the Higgs potential to depend on explicit sources of the symmetry breaking and to vanish if the latter are sent to zero, namely if the new physics sector decouples from the SM. Partial compositeness \cite{Kaplan:1991dc} is typically a key ingredient of pNGB Higgs models to transmit the electroweak symmetry breaking (EWSB) from the strong sector to the elementary sector, namely to the SM fermions, or at the very least to the top quark. For reviews see \cite{Contino:2010rs,Bellazzini:2014yua,Panico:2015jxa} and references therein. Partial compositeness implies a linear mixing of each chirality of the SM fermions with an operator of the strong sector with matching quantum numbers: this usually implies the presence of vector-like fermionic partners belonging to the realm of the resonances, i.e. bound states of the strongly coupled sector. Since partial compositeness is effectively a see-saw mechanism \cite{Contino:2006nn}, large SM masses are favored for large ratios of mixing over masses: this implies that the resonance coupling to the top needs to be light, if we want to keep mixing couplings perturbative and small perturbations of the strong sector.
Moreover these resonances contribute to the Higgs potential via the linear mixings and, under some assumptions~\cite{Marzocca:2012zn}, help taming its sensitivity to high energy scale physics~\footnote{We should mention that there are models where the stability of the Higgs potential does not rely on top partner loops, as in~\cite{Galloway:2010bp,Cacciapaglia:2014uja,Arbey:2015exa,Ma:2015gra}.}. Light top partners are generically favored by considerations on the Higgs potential \cite{Marzocca:2012zn,Matsedonskyi:2012ym,Pomarol:2012qf}, although there are ways to evade this conclusion \cite{Panico:2012uw,Carmona:2014iwa}. Light un-colored fermions are also needed in some composite Twin Higgs models \cite{Barbieri:2015lqa,Low:2015nqa}. Finally the possibility of light top partners is interesting because it can be directly tested at the LHC: Run I established limits around $800$ GeV on their masses \cite{Chatrchyan:2013wfa,Aad:2015kqa,Khachatryan:2015oba}, while Run II will explore masses up to $1.4$ TeV \cite{Matsedonskyi:2014mna,Matsedonskyi:2015dns} (or at most up to $2$ TeV \cite{Backovic:2014uma}) and heavier fermions are certainly outside the reach of the machine.

The masses of fermionic resonances in a generic strongly coupled theory are expected to be at the same scale as the mass of vector resonances, i.e. in the multi-TeV range. For scalar resonances, appearing as pNGBs, a shift symmetry can bear the responsibility of their small mass, however no such symmetry is present for fermions. There exist extra dimensional constructions where this single scale degeneracy is relieved and Kaluza-Klein fermionic states are lighter than others, as in \cite{Agashe:2003zs,Agashe:2004ci,Agashe:2004bm,Carena:2006bn,Contino:2006qr}.
Following the AdS/CFT correspondence, such theories are dual to 4 dimensional (4D) strongly coupled Conformal Field Theories and the light states interpreted as zero modes of spin-1/2 operators~\cite{Contino:2004vy} (see also~\cite{Cacciapaglia:2008bi} for a supersymmetric analysis): the presence of such states crucially depends on the dimension of the operator in the Conformal theory.
In this paper we want to focus on the possibility that the top partners emerge as bound states of an underlying 4D confining dynamics (not necessarily Conformal), which only contains matter fermions: in such case, it is non-trivial to obtain feasible operators \cite{Ferretti:2013kya}. A simple way-out that allows to protect fermion masses could be to introduce supersymmetry \cite{Caracciolo:2012je,Marzocca:2013fza,Parolini:2014rza}, at the price of reintroducing fundamental scalars \footnote{A model of composite vector-like fermions, formed by a fermion and a scalar, which become light for the binding coupling close to a critical value is studied in \cite{Dobrescu:2014zla}.}.

In this letter we want to stick to classes of models with purely fermionic components, and propose the possibility that light composite fermions may be present in the spectrum due to the 't~Hooft anomaly matching \cite{'tHooft:1979bh}.
This possibility, and related ideas, has been widely employed in the 80s ~\cite{Dimopoulos:1980hn,Banks:1981xr} in the quest for light composite fermions that may play the role of the SM fermions: examples can be found in \cite{Barbieri:1980aq,Dimopoulos:1981xc,Preskill:1981sr,Casalbuoni:1982ei}.
The template model we will consider is based on a strong sector based on a hypercolor (HC) gauge group $G_{HC}$ with one, or more, species of fermions transforming under different representations of the HC group.
The global symmetry $\mathcal{G}$ of the model is thus determined by the number of species and their multiplicity.
We also assume that the confinement of the HC group generates one or more fermion condensates which are non-vanishing in the vacuum of the theory, thus breaking $\mathcal{G}$ to a subgroup $\mathcal{H}$.
The 't~Hooft anomaly matching is based on the fact that $\mathcal{H}$ may suffer from global anomalies in the underlying theory, which depend on the details of the underlying dynamics.
In the confined phase, the value of the anomaly should be matched by the presence of massless composite fermions, transforming under suitable representations of $\mathcal{H}$.
The matching is highly non trivial as the representations of $\mathcal{H}$ in the spectrum are constrained by the structure of the fermionic content in the underlying theory, while the anomaly in the un-confined phase depends on the number of hypercolors.
If on the other hand no solution is found the breaking pattern $\mathcal{G} \to \mathcal{H}$ cannot be realised, and additional condensates must be turned on.
Further conditions may apply if one considers the decoupling limit of the underlying fermions, however this relies on the absence of phase transitions and, following \cite{Preskill:1981sr}, we will not consider it here. 
Once the low energy theory contains 't~Hooft composite fermions a mass term for them can then be generated by adding an explicit breaking of $\mathcal{H}$, for instance in the form of a mass term $m_\chi$ for the fundamental fermions: the top partner masses must therefore vanish for vanishing $m_\chi$, and their value can be made parametrically smaller than the mass of other resonances. Hence, such situation generates technically natural light masses.

We should stress, however, that finding a solution for the 't~Hooft matching does not imply that such spectrum is realised: the vacuum is dynamically chosen by the strong sector, and it can only be determined by non-perturbative techniques, like on the Lattice, or in suitable limits of the theory, such as large $N_{HC}$ expansion. Furthermore, the presence of an explicit, albeit small, breaking of the global symmetry may destabilise the vacuum.

In the following we illustrate the mechanism for few realistic models: in a case where it works in section \ref{sec: modello di gherghetta} and in two models where it is not applicable in section \ref{sec: altri modelli}, while in section \ref{sec: conclusioni} we conclude. We leave for a future work a more systematic study, characterizing theories in terms of the possibility of matching global anomalies.

\section{The minimal model SU$(N_Q)\to$Sp$(N_Q)$}\label{sec: modello di gherghetta}

The first model under consideration consists of a $G_{HC} = \Sp(2N_c)$ gauge theory with two species of chiral fermions: $Q$ in the fundamental and $\chi$ in the two-index anti-symmetric, where the absence of the Witten anomaly requires even $N_Q$. The largest global symmetry is therefore $\mathcal{G} = \SU(N_Q) \times \SU(N_\chi) \times \U(1)$. The model is summarized in Table~\ref{tab:mod1} and the minimal case of CHM corresponds to $N_Q = 4$ and $N_\chi = 6$ \cite{Barnard:2013zea}.
\begin{table}[t]
\begin{center}
\begin{tabular}{|c|c|ccc|}
\hline
& Sp$(2N_c)$& SU$(N_Q)$ & SU$(N_{\chi})$ & U(1) \\\hline\hline
$Q$ & ${\tiny{\yng(1)}}$ &$ N_Q$&$1$ & $1$ \\
$\chi$ & ${\tiny{\yng(1,1)}}$&1&$N_\chi$&  $- \frac{N_Q}{2 N_\chi (N_c-1)}$ \\  
\hline
\end{tabular} 
\caption{Fermionic field content of the first model. }
\label{tab:mod1}
\end{center}
\end{table}

In the confined phase, there exist potentially two fermion condensates that spontaneously break the global symmetries when assuming a non-zero value on the vacuum: $\langle Q Q \rangle$ breaks $\SU(N_Q) \times \U(1) \to \Sp (N_Q)$ thus leading to a pNGB Higgs in the spectrum; $\langle \chi \chi \rangle$ would break $\SU(N_\chi) \times \U(1) \to \SO(N_\chi)$, where the QCD color $\SU(3)_c \subset \SO(N_\chi)$. In the following we will consider two phases that lead to a potentially interesting pNGB Higgs model.

\paragraph{Phase $\langle QQ \rangle \neq 0$, $\langle \chi \chi \rangle \neq 0$.} In this case, the unbroken global symmetry is $\mathcal{H} =\Sp(N_Q) \times \SO(N_\chi)$, which has no global anomalies. Therefore, all composite fermions are heavy and there is no symmetry reason why some of them should be parametrically lighter than the other resonances. On the other hand, the spontaneous breaking of $\SU(N_\chi)$ implies the presence of light colored scalars \cite{Cacciapaglia:2015eqa}.

\paragraph{Phase $\langle QQ \rangle \neq 0$, $\langle \chi \chi \rangle = 0$.} In this phase, $\mathcal{H} = \Sp(N_Q) \times \SU(N_\chi)$. As the global symmetry containing the QCD color is unbroken, no light colored pNGBs will be present in the spectrum. In the underlying theory, $\SU(N_\chi)$ has a global anomaly proportional to
\begin{equation}
A_{\SU(N_\chi)^3} =  \mbox{dim} ({\tiny{\yng(1,1)}}) = (2 N_c+1) (N_c-1)\,.
\end{equation}
This anomaly should be matched by the composite fermions in the confined phase: considering the lowest dimensional operators, we have three-fermion states $QQ\chi$ ($\bar{Q} \bar{Q} \chi$) and $Q \bar{Q} \bar{\chi}$, where all states have the same chirality. For $N_c > 2$, bound states $\chi \chi \chi$ and $\chi \bar{\chi} \bar{\chi}$ are also possible, however they do not couple to the Higgs in the $\SU(N_Q)/\Sp(N_Q)$ coset and are therefore not suitable to be top partners. Bound states with larger number of fermionic components are also possible, however we will not consider them here as they are likely to quickly decay into lower dimensional states and may thus be highly unstable.
In this model, all top partners transform as either the fundamental ($\bf F$) or anti-fundamental ($\bf \bar{F}$) of $\SU(N_\chi)$, thus their contribution to the ${\SU(N_\chi)}^3$ anomaly is simply given by the multiplicity of the representation under $\Sp(N_Q)$: as $Q$ and $\bar{Q}$ transform as the fundamental of $\Sp(N_Q)$, all bound states will contain a singlet $\bf 1$, a two-index symmetric $\bf S$ and a two-index anti-symmetric $\bf A$, as shown in Table~\ref{tab:reps}.
\begin{table}[t]
\begin{center}
\begin{tabular}{|c|c|c|c|c|}
\hline
   & \footnotesize{$\SU(N_Q)\times\SU(N_\chi)$} & \footnotesize{$\Sp(N_Q)\times\SU(N_\chi)$} & $d_{\Sp(N_Q)}$ \\
\hline \hline
\multirow{3}{*}{$\begin{array}{c}\chi Q Q \\ \chi \bar{Q} \bar{Q}\end{array}$} &  \multirow{2}{*}{$(\bf{A}, \bf{F})$} & $(\bf{1}, \bf{F})$ & $1$ \\
                  &                              & $(\bf{A}, \bf{F})$ & $\frac{N_Q (N_Q-1)}{2} -1$ \\
                  &   $(\bf{S}, \bf{F})$ & $(\bf{S}, \bf{F})$ & $\frac{N_Q(N_Q+1)}{2}$ \\
\hline
\multirow{3}{*}{$\bar{\chi} \bar{Q} Q$} &  $(\bf{1}, \bf{\bar F})$ & $(\bf{1}, \bf{\bar F})$ & $1$ \\
                  &    \multirow{2}{*}{$(\bf{Adj}, \bf{\bar F})$}    & $(\bf{A}, \bf{\bar F})$ & $\frac{N_Q (N_Q-1)}{2} -1$ \\
                  &                                                           & $(\bf{S}, \bf{\bar F})$  & $\frac{N_Q(N_Q+1)}{2}$ \\
\hline
\end{tabular} 
\caption{Three fermions bound states of the model with group properties with respect to the global flavor group and the unbroken subgroups.}
\label{tab:reps}
\end{center}
\end{table}
The anomaly matching condition can thus be simply expressed as
\begin{multline}
n_{\bf 1} + \left( \frac{N_Q (N_Q-1)}{2} -1 \right)\ n_{\bf A} + \frac{N_Q(N_Q+1)}{2}\ n_{\bf S} = \\ (2 N_c+1) (N_c-1)\,,
\end{multline}
which gives a non-trivial relation between the number of flavors of $Q$ and number of hypercolors. In the above equation, $n_{\bf X}$ is the difference of the number of fundamental and anti-fundamentals of $\SU(N_\chi)$ in the representation $\bf X$ of $\Sp(N_Q)$. Due to the presence of singlets of $\Sp(N_Q)$, the above condition always has a trivial solution when the multiplicity of hypercolors is matched by the number of massless singlets.

Very attractive solutions can be achieved if a relation between the number of $Q$-flavors and the number of hypercolors is present: for instance, a single anti-symmetric of $\Sp(N_Q)$ is sufficient if $N_Q = 2 N_c$. In the minimal model with $G_{HC} =\Sp(4)$ this singles out top partners in the  ${\bf A}={\bf 5}$ of $\Sp(4)\simeq\SO(5)$.
Interestingly, there are no solutions where the only massless fermion is an $\bf S$, nor when only an $\bf A$ and a singlet are present: these cases would correspond to a single complete $\SU(N_Q)$ representation. In the minimal model, we also found a simple solution containing one $\bf S = 10$ in the fundamental of $\SU(6)$, and one $\bf A = 5$ in the anti-fundamental, corresponding to $n_{\bf S} = 1 = - n_{\bf A}$ and $n_{\bf 1} = 0$.

The global $\SU(N_\chi)$ can be explicitly broken to $\SO(N_\chi)$ by giving a gauge invariant mass to the $\chi$'s, so that the massless fermions will acquire a mass that scales with $m_\chi$.
Note also that the global $\Sp(N_Q)$ potentially suffers from a global Witten anomaly \cite{Bhansali:1992yg}, however it identically vanishes in this model.

\section{Other models}\label{sec: altri modelli}
In this section we discuss other models that recently appeared in the literature, in particular we check whether this mechanism to protect the mass of top partners can work in the models considered in \cite{Ferretti:2014qta} and \cite{Vecchi:2015fma}. Unfortunately this is not the case, meaning that 't~Hooft anomalies cannot be matched in one case and identically vanish in the other.

The model in \cite{Ferretti:2014qta} consists of a $G_{HC} = \SU(4)$ gauge theory and three species of left-handed fermions with quantum numbers as in Table~\ref{tab:ferretti}.
\begin{table}[t]
\begin{center}
\begin{tabular}{|c|c|cccc|}
\hline
& SU$(4)$& SU$(5)$ & $\SU(3)\times\SU(3)'$ & ${\U(1)}_X$&U(1)' \\\hline\hline
$Q$ & $\bf6$ &$\bf5$&$({\bf1},{\bf 1})$ & 0$$&-1\\
$\chi$ & $\bf4$&\bf1&$({\bf3},{\bf1})$&-1/3$$&5/3\\
$\tilde{\chi}$ & $\bf \overline{4}$&\bf1&$({\bf1},{\bf\overline3})$&$1/3$&5/3\\\hline
\end{tabular} 
\caption{Fermionic field content of the theory of \cite{Ferretti:2014qta}, and transformation under the global symmetry $\SU(5) \times \SU(3) \times \SU(3)' \times \U(1)^2$.}
\label{tab:ferretti}
\end{center}
\end{table}
The SM color is identified as the diagonal $\SU(3)$ in the $\SU(3)\times\SU(3)'$. The pNGB Higgs boson is coming from $\langle QQ\rangle \neq 0$, that spontaneously breaks  $\SU(5) \times \U(1)'$ to $\SO(5)$. The second condensate that may form is $\langle\tilde\chi\chi\rangle$ that, if non-vanishing on the vacuum, would break $\SU(3) \times \SU(3)' \to \SU(3)_c$. In the broken phase, no global anomalies are present, as the unbroken $\SU(3)$ corresponds to the gauged color.
However, in the vacuum  $\langle\tilde\chi\chi\rangle=0$, we need to match $A_{({\SU(3)}^3)}$ and $A_{({\SU(3)'}^3)}$ anomalies. In the underlying theory $A_{({\SU(3)}^3)}=-A_{({\SU(3)'}^3)}=4$. All three-fermion bound states of the low energy theory \cite{Ferretti:2014qta} are of the form $\chi Q\chi$, and they transform as fundamental or two-index antisymmetric of the $\SU(3)$ flavor symmetries, thus a matching of the anomaly seems possible. However, as they all contain a single $Q$, they all come with a multiplicity of $5$, which means that in the low energy theory we get an anomaly coefficient multiple of $5$ and not $4$. We leave as an open question the role of bound states with more than three fermions, as $QQQ\chi\chi$, which in principle can transform as $\bf3$ or $\bf\bar 3$ of one of the two global $\SU(3)$.
This conclusion can be generalised to models with an arbitrary number of fermions, as the presence of a pNGB Higgs in the spectrum always requires that $N_Q \geq 5$.

The second model we analyse was proposed in \cite{Vecchi:2015fma}: it is based on a $G_{HC} = \SU(3)$ gauge theory with seven vector-like fermions. 
The SM gauge interactions are embedded in the diagonal $\SU(7)$ flavor symmetry, while top partners arise as three-fermion bound states, like baryons in QCD. A pNGB Higgs can be obtained in a similar way as in the more minimal coset $\SU(4)^2 \to \SU(4)$ \cite{Ma:2015gra}.
Setting aside the SM gauge interactions, it is a theory with a $\SU(7)\times\SU(7)\times\U(1)$ global symmetry, broken in the vacuum by a condensate to the diagonal $\SU(7)$, and this does not have anomalies to be matched since it is a vector-like theory. We believe unlikely that the strong dynamics breaks only partially the flavor symmetry to the diagonal $\SU(n)$ leaving a chiral $\SU(7-n)\times\SU(7-n)$ unbroken, because strong effects are flavor blind. Thus in this theory no global anomaly can enforce composite fermions to be massless. 

\section{Conclusions}\label{sec: conclusioni}
In the context of Composite pNGB Higgs models with top partial compositeness we studied conditions under which the low energy theory might contain light top partners,  lighter than the estimate based on naive dimensional analysis. The 't~Hooft anomaly matching condition is proposed as a mechanism to force some fermions to be massless, or parametrically light once an explicit soft breaking of the global symmetry is introduced. We find this idea attractive because we are not aware of any other mechanism with the same effect, in four dimensions without elementary scalars, and because light top partners play an important role in model building and in new physics phenomenology at the LHC.

We examined in details specific models based on an UV free gauge theory of interacting fermions, proposed in the literature as underlying theories of pNGB Higgs with top partners, where we find no obstruction for this mechanism, namely the model possesses a global symmetry whose anomaly can be matched by composite fermions with the quantum numbers of a top partner candidate.
For this mechanism to work, we need at least two species of fundamental fermions since for one of them a global symmetry should survive while the other has to provide a condensate. 
This mechanism, however, does not apply to any UV construction: out of the three models under scrutiny, we found that solutions of the 't~Hooft anomalies are only possible in one scenario, provided that one of the two fermion condensates does not occur.
In the other two cases, either there are no solutions to the anomaly matching, or the anomalies vanish.
We believe this mechanism can thus be a useful criterion to select interesting models to be further studied, although we leave a thorough classification to a future work. The main question that remains open is about the vacuum that the theory chose to live in, and finding an answer requires a study of the model on the Lattice, or by means of other non-perturbative techniques.

\section*{Acknowledgements}
We are particularly grateful to Aldo Deandrea and Hugo Ser\^{o}dio for fruitful discussion, and also to Haiying Cai, Thomas Flacke, Seung Joon Lee, Micka\"el Lespinasse and Jeong Han Kim.  AP acknowledges Luca Vecchi for interesting conversations. 
GC acknowledges partial support from the Labex-LIO (Lyon Institute of Origins) under grant ANR-10-LABX-66 and FRAMA (FR3127, F\'ed\'eration de Recherche ``Andr\'e Marie Amp\`ere").

\end{document}